\documentclass[aps,prd,superscriptaddress,amsmath,amsfont,amssymb,preprint,nofootinbib]{revtex4}
\pdfoutput=1
\allowdisplaybreaks
\usepackage{graphicx}
\usepackage{ bbold }
\usepackage{subfigure}
\usepackage{slashed}
\usepackage{mathtools}
\usepackage[english]{babel}
\newcommand{\op}{{\cal O}}

\newcommand{\C}{{\cal C}}
\newcommand{\Q}{{\cal Q}}
\newcommand{\todo}[1]{{\color{red} \ifmmode\else[todo]\fi #1}}
\usepackage[usenames,dvipsnames]{xcolor}
     \definecolor{hgreen}{rgb}{0,.3,0}
     \definecolor{hred}{rgb}{.3,0,0}
     \definecolor{hblue}{rgb}{0,0,.3}
     \definecolor{LightGray}{gray}{0.95}
\usepackage[backref=page,
            colorlinks=true,
            linkcolor=hblue,
            citecolor=hgreen,
            filecolor=hblue,
            urlcolor=hred]{hyperref}
\usepackage{fancyvrb}
\usepackage{upquote}

\renewcommand*{\backref}[1]{}

\newcommand{\beq}{\begin{equation} }
\newcommand{\eeq}{\end{equation}} 
\newcommand{\bi}{\begin{itemize} }
\newcommand{\ei}{\end{itemize} }

\definecolor{Red}{rgb}{1.,0.,0.}
\definecolor{Grn}{rgb}{0.,0.75,0.}
\definecolor{Blu}{rgb}{0.,0.,1.}

\def\arraystretch{0.85}

\definecolor{pan624}{rgb}{0.482,0.635,0.588} 
\definecolor{pan576}{rgb}{0.412,0.569,0.231} 
\definecolor{pan129}{rgb}{0.961,0.812,0.278}
\definecolor{pan5405}{rgb}{0,0.129,0.278} 
\definecolor{shadecolor}{rgb}{0.482,0.635,0.588}
\definecolor{mygray}{HTML}{666666}
\definecolor{x11steelblue}{HTML}{4682B4}
\definecolor{x11firebrick}{HTML}{B22222}
\definecolor{x11forestgreen}{HTML}{228B22}

\newcommand{\nfiii}{{\tt 3Flavor}}
\newcommand{\nfiv}{{\tt 4Flavor}}
\newcommand{\nfv}{{\tt 5Flavor}}
\newcommand{\nr}{{\tt NR}}
\newcommand{\ddm}{\texttt{DirectDM}}
\newcommand{\mma}{{\texttt{Mathematica}}}
\usepackage{tikz}
\tikzstyle{every picture}+=[remember picture]
\usetikzlibrary{shapes.geometric}
\usetikzlibrary{calc}
\usetikzlibrary{decorations.markings}
\usetikzlibrary{decorations.text}
\usetikzlibrary{patterns}
\usetikzlibrary{backgrounds}
\usetikzlibrary{positioning}
\tikzstyle arrowstyle=[scale=2]
\tikzstyle directed=[postaction={decorate,decoration={markings,
		mark=at position 0.6 with {\arrow[arrowstyle]{>}}}}]
\tikzstyle rarrow=[postaction={decorate,decoration={markings,
		mark=at position 0.999 with {\arrow[arrowstyle]{>}}}}]
		
\graphicspath{{./figs/}}

\begin{document}

\title{{\tt DirectDM}: a tool for dark matter direct detection}

\def\Cincy{Department of Physics, University of Cincinnati, Cincinnati, Ohio 45221,USA}
\def\UCSD{Department of Physics, University of California-San Diego, La Jolla, CA 92093, USA}
\def\Mainz{PRISMA Cluster of Excellence \& Mainz Institute for Theoretical
Physics, Johannes Gutenberg University, 55099 Mainz, Germany}
\def\TUD{Fakult\"at f\"ur Physik, TU Dortmund, D-44221 Dortmund, Germany} 
\def\CERN{CERN, Theory Division, CH-1211 Geneva 23, Switzerland}
\def\Oxford{Rudolf Peierls Centre for Theoretical Physics, University of Oxford OX1 3NP Oxford, United Kingdom}

\author{\textbf{Fady Bishara}}
\email{fady.bishara AT physics.ox.ac.uk}
\affiliation{\Oxford}

\author{\textbf{Joachim Brod}}
\email{joachim.brod AT tu-dortmund.de}
\affiliation{\TUD}

\author{\textbf{Benjamin Grinstein}}
\email{bgrinstein AT ucsd.edu}
\affiliation{\UCSD}

\author{\textbf{Jure Zupan}} 
\email{zupanje AT ucmail.uc.edu}
\affiliation{\Cincy}
\affiliation{\CERN}

\date{\today}

\begin{abstract}
We provide a {\tt Mathematica} package, {\tt DirectDM}, that takes as
input the Wilson coefficients of the relativistic effective theory
describing the interactions of dark matter with quarks, gluons and
photons, and matches it onto an effective theory describing the
interactions of dark matter with neutrons and protons. The
nonperturbative matching is performed at leading order in a chiral
expansion. The one-loop QCD and QED renormalization-group evolution
from the electroweak scale down to the hadronic scale, as well as
finite corrections at the heavy quark thresholds are taken into
account. We also provide an interface with the package {\tt
  DMFormFactor} so that, starting from the relativistic effective
theory, one can directly obtain the event rates for direct detection
experiments.
\end{abstract}

\pacs{--pacs--}

\preprint{DO-TH 17/11}
\preprint{OUTP-17-08P}

\maketitle
\tableofcontents


\section{Introduction}
\label{sec:Intro}
Dark Matter (DM) scattering on nuclei in direct detection experiments
is naturally described by an effective field theory (EFT)
\cite{Fan:2010gt, Fitzpatrick:2012ix, Fitzpatrick:2012ib,
  Anand:2013yka, Cirigliano:2012pq, DelNobile:2013sia,
  Barello:2014uda, Hill:2014yxa, Hoferichter:2015ipa, Catena:2014uqa,
  Kopp:2009qt, Hill:2013hoa, Hill:2011be, Hoferichter:2016nvd,
  Kurylov:2003ra, Pospelov:2000bq, Bagnasco:1993st, Bishara:2016hek,
  Bishara:2017pfq}, since the typical momentum exchange for the DM
scattering on a nucleus, $q\lesssim 100$ MeV, is much smaller than the
mediator mass in most DM models. In Refs.~\cite{Bishara:2016hek,
  Bishara:2017pfq} we presented the analytic expressions for the
matching, within chiral perturbation theory (ChPT), between two EFTs
describing the interactions of DM with the standard model (SM). The
first, ``relativistic EFT'', comprises the partonic interactions of DM
with quarks, gluons, and photons, while the second, ``nonrelativistic
EFT'', describes the nonrelativistic interactions of DM with nucleons
-- neutrons and protons~\cite{Fitzpatrick:2012ix, Fitzpatrick:2012ib,
  Anand:2013yka}. Here, we introduce the {\tt Mathematica} package
\ddm{} which takes as input the Wilson coefficients of the
relativistic operators and performs the nonperturbative matching onto
the nonrelativistic EFT. An interface with {\tt DMFormFactor}
\cite{Anand:2013yka} is provided so that, starting from the
relativistic theory, one can obtain directly the event rates in the
experiment. The \ddm{} code can be downloaded from
\begin{center}
\url{https://directdm.github.io}
\end{center}

This paper is organized as follows. In Section~\ref{sec:basis} we fix
our notation and introduce the bases of both the relativistic and
nonrelativistic theories. We also include a short discussion of the
renormalization-group (RG) evolution of the Wilson coefficients.
Section~\ref{sec:program} contains the manual for the {\tt
  DMFormFactor} package. We conclude in
Section~\ref{sec:conclusion}. Appendix~\ref{app:majorana-and-real-scalar}
gives the operator bases for Majorana and real scalar DM, while
Appendix \ref{sec:tait} contains the translation to the operator bases
of Ref.~\cite{Goodman:2010qn}.

\section{Operator basis and renormalization-group evolution}
\label{sec:basis}
\subsection{Fermionic dark matter}
The starting point is the interaction Lagrangian between fermionic DM
and the SM, which is given in terms of higher dimension operators,
\begin{equation}\label{eq:lightDM:Lnf5}
{\cal L}_\chi=\sum_{a,d}
\hat \C_{a}^{(d)} {\cal Q}_a^{(d)}, 
\qquad {\rm where}\quad 
\hat \C_{a}^{(d)}=\frac{\C_{a}^{(d)}}{\Lambda^{d-4}}\,.
\end{equation}
Here, the $\C_{a}^{(d)}$ are dimensionless Wilson coefficients, while
$\Lambda$ can be identified with the mediator mass. The Wilson
coefficients depend on the renormalization scale $\mu$ (see also the
discussion below and in Section~\ref{sec:RG}). The sum runs over the
mass dimension of the operators, $d=5,6,7$, as well as the index $a$
of the individual operators. We keep all dimension-five and
dimension-six operators, all dimension-seven operators coupling DM to
gluons, and the most relevant subset of dimension-seven operators that
couple DM to quarks (i.e., we do not keep the operators that are
additional suppressed by derivatives -- see~\cite{future:BGTZ} for the
complete basis).
  
We start with DM that is a Dirac fermion. The operator basis is the
same as in \cite{Bishara:2017pfq}. There are two dimension-five
operators,
\begin{equation}
\label{eq:dim5:nf5:Q1Q2:light}
{\cal Q}_{1}^{(5)} = \frac{e}{8 \pi^2} (\bar \chi \sigma^{\mu\nu}\chi)
 F_{\mu\nu} \,, \qquad {\cal Q}_2^{(5)} = \frac{e }{8 \pi^2} (\bar
\chi \sigma^{\mu\nu} i\gamma_5 \chi) F_{\mu\nu} \,,
\end{equation}
where $F_{\mu\nu}$ is the electromagnetic field strength tensor. The
magnetic dipole operator $\Q_1^{(5)}$ is CP even, while the electric
dipole operator $\Q_2^{(5)}$ is CP odd. The dimension-six operators
are
\begin{align}
{\cal Q}_{1,f}^{(6)} & = (\bar \chi \gamma_\mu \chi) (\bar f \gamma^\mu f)\,,
 &{\cal Q}_{2,f}^{(6)} &= (\bar \chi\gamma_\mu\gamma_5 \chi)(\bar f \gamma^\mu f)\,, \label{eq:dim6:Q1Q2:light}
  \\ 
{\cal Q}_{3,f}^{(6)} & = (\bar \chi \gamma_\mu \chi)(\bar f \gamma^\mu \gamma_5 f)\,,
  & {\cal Q}_{4,f}^{(6)}& = (\bar
\chi\gamma_\mu\gamma_5 \chi)(\bar f \gamma^\mu \gamma_5 f)\,.\label{eq:dim6:Q3Q4:light}
\end{align}
The dimension-seven operators that we keep are 
\begin{align}
{\cal Q}_1^{(7)} & = \frac{\alpha_s}{12\pi} (\bar \chi \chi)
 G^{a\mu\nu}G_{\mu\nu}^a\,, 
 & {\cal Q}_2^{(7)} &= \frac{\alpha_s}{12\pi} (\bar \chi i\gamma_5 \chi) G^{a\mu\nu}G_{\mu\nu}^a\,,\label{eq:dim7:Q1Q2:light}
 \\
{\cal Q}_3^{(7)} & = \frac{\alpha_s}{8\pi} (\bar \chi \chi) G^{a\mu\nu}\widetilde
 G_{\mu\nu}^a\,, 
& {\cal Q}_4^{(7)}& = \frac{\alpha_s}{8\pi}
(\bar \chi i \gamma_5 \chi) G^{a\mu\nu}\widetilde G_{\mu\nu}^a \,, \label{eq:dim7:Q3Q4:light}
\\
{\cal Q}_{5,f}^{(7)} & = m_f (\bar \chi \chi)( \bar f f)\,, 
&{\cal
  Q}_{6,f}^{(7)} &= m_f (\bar \chi i \gamma_5 \chi)( \bar f f)\,,\label{eq:dim7:Q5Q6:light}
  \\
{\cal Q}_{7,f}^{(7)} & = m_f (\bar \chi \chi) (\bar f i \gamma_5 f)\,, 
&{\cal Q}_{8,f}^{(7)} & = m_f (\bar \chi i \gamma_5 \chi)(\bar f i \gamma_5
f)\,, \label{eq:dim7:Q7Q8:light}  
 \\
{\cal Q}_{9,f}^{(7)} & = m_f (\bar \chi \sigma^{\mu\nu} \chi) (\bar f \sigma_{\mu\nu} f)\,, 
&{\cal Q}_{10,f}^{(7)} & = m_f (\bar \chi  i \sigma^{\mu\nu} \gamma_5 \chi)(\bar f \sigma_{\mu\nu}
f)\,. \label{eq:dim7:Q9Q10:light} 
\end{align}
Here $G_{\mu\nu}^a$ is the QCD field strength tensor, while
$\widetilde G_{\mu\nu} = \tfrac{1}{2} \varepsilon_{\mu\nu\rho\sigma}
G^{\rho\sigma}$ is its dual, and $a=1,\dots,8$ are the adjoint color
indices. Moreover, $\chi$ denotes the DM fields and $f$ the SM fermion
fields\footnote{Although we are primarily interested in the hadronic
  effects, we keep the SM leptons explicit in our definitions, since
  the leptonic operators mix into the hadronic ones via QED penguins,
  see Sec.~\ref{sec:RG}.}. The operators can be specified in the
three-flavor ($f=u,d,s,e,\mu,\tau$), four-flavor
($f=u,d,s,c,e,\mu,\tau$), and five-flavor scheme
($f=u,d,s,c,b,e,\mu,\tau$). The initial conditions for the Wilson
coefficients have then to be specified at the scale $\mu_c = 2\,$GeV
(three-flavor), $\mu_b = m_b(m_b)=4.18\,$GeV (four-flavor), or $\mu_Z
= M_Z = 91.1876\,$GeV (five-flavor), respectively. The scheme is set
in the code by choosing one of the options \nfiii, \nfiv, or \nfv{},
see Section~\ref{sec:program}. In the first case (three-flavor
scheme), the Wilson coefficients are directly matched to the nuclear
effective theory, see below, while in the latter two cases the code by
default performs the QCD and QED RG running down to the hadronic scale
$\mu_h = 2\,$GeV, with the subsequent matching to the nuclear theory.
For Majorana DM, the operators $\Q_{1,2}^{(5)}$, $\Q_{1,f}^{(6)}$,
$\Q_{3,f}^{(6)}$, $\Q_{9,f}^{(7)}$, and $\Q_{10,f}^{(7)}$ vanish,
while the definitions of all the other operators include an additional
factor of $1/2$, see Appendix~\ref{app:majorana-and-real-scalar}.
Frequently, the operator basis of Ref.~\cite{Goodman:2010qn} is used
in phenomenological analyses. We provide the translation to our basis
in App.~\ref{sec:tait}.

\begin{figure}
\includegraphics[scale=0.5]{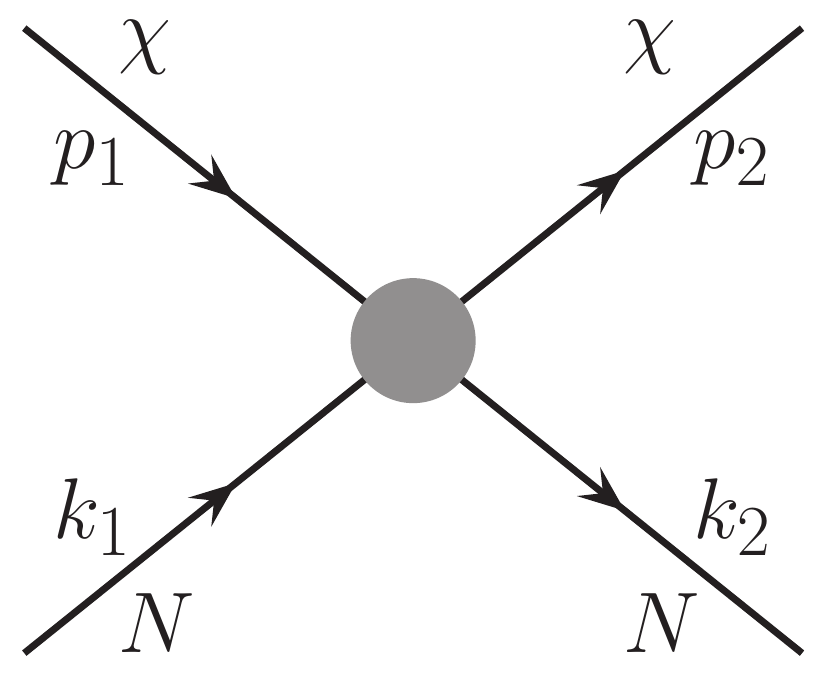}~~~~~~~~~~
\caption{The kinematics of DM scattering on nucleons,
  $\chi(p_1)N(k_1)\to \chi(p_2) N(k_2)$.  }
         \label{fig:scattering_kin}
\end{figure}

The \ddm{} code provides the matching between the EFT coupling DM to
quarks, gluons and photons, given in Eq.~\eqref{eq:lightDM:Lnf5}, to
the EFT where DM interacts with nonrelativistic nucleons, given by the
Lagrangian
\begin{equation}\label{eq:LNR}
{\cal L}_{\rm NR}=\sum_{i,N} c_i^N(q^2) \op_i^N\,.
\end{equation}
We implement the expressions for the coefficients $c_i^N(q^2)$ to
leading order (LO) in the chiral expansion, i.e., to LO in an
expansion in the momentum transfer $q/(4\pi f)$. At this order, ${\cal
  L}_{\rm NR}$ contains two momentum-independent nonrelativistic
operators,
\begin{align}
\label{eq:O1pO4p}
{\mathcal O}_1^N&= \mathbb{1}_\chi \mathbb{1}_N\,,
&{\mathcal O}_4^N&= \vec S_\chi \cdot \vec S_N \,,
\end{align}
and a set of momentum-dependent operators,
\begin{align}
\label{eq:O5pO6p}
{\mathcal O}_5^N&= \vec S_\chi \cdot \Big(\vec v_\perp \times \frac{i\vec q}{m_N} \Big) \, \mathbb{1}_N \,,
&{\mathcal O}_6^N&= \Big(\vec S_\chi \cdot \frac{\vec q}{m_N}\Big) \, \Big(\vec S_N \cdot \frac{\vec q}{m_N}\Big),
\\
\label{eq:O7pO8p}
{\mathcal O}_7^N&= \mathbb{1}_\chi \, \big( \vec S_N \cdot \vec v_\perp \big)\,,
&{\mathcal O}_8^N&= \big( \vec S_\chi \cdot \vec v_\perp \big) \, \mathbb{1}_N\,,
\\
\label{eq:O9pO10p}
{\mathcal O}_9^N&= \vec S_\chi \cdot \Big(\frac{i\vec q}{m_N} \times \vec S_N \Big)\,,
&{\mathcal O}_{10}^N&= - \mathbb{1}_\chi \, \Big(\vec S_N \cdot \frac{i\vec q}{m_N} \Big)\,,
\\
\label{eq:O11pO12p}
{\mathcal O}_{11}^N&= - \Big(\vec S_\chi \cdot \frac{i\vec q}{m_N} \Big) \, \mathbb{1}_N \,,
&{\mathcal O}_{12}^N&= \vec S_\chi \cdot \Big( \vec S_N \times \vec v_\perp \Big) \,,
\end{align}
with $N=p,n$. These operators coincide with the ones defined
in~\cite{Anand:2013yka}, while our definition of the momentum exchange
differs by a minus sign with respect to the convention used
in~\cite{Anand:2013yka}, so that (cf. Fig. \ref{fig:scattering_kin})
\begin{equation}
\vec q = \vec k_2-\vec k_1=\vec p_1 -\vec p_2\,, \qquad \vec v_\perp=
\frac{\vec p_1+\vec p_2}{2{m_\chi}} - \frac{\vec k_1+\vec
  k_2}{2{m_N}}\,.
\end{equation}

The $q^2$-dependent coefficients $c_i^N$ in Eq.~\eqref{eq:LNR} are given
by~\cite{Bishara:2017pfq} 
\begin{align}
\label{eq:cNR1}
c_{1}^p &=-\frac{\alpha}{2\pi m_\chi} Q_p \hat \C_1^{(5)}+ \sum_q \Big( F_{1}^{q/p} \, \hat
\C_{1,q}^{(6)} + F_{S}^{q/p} \, \hat \C_{5,q}^{(7)} \Big) + F_G^p \,
\hat \C_{1}^{(7)} 
\\
&\qquad - \frac{\vec q^{\,\,2}}{2m_\chi m_N} \sum_q \big( 
F_{T,0}^{q/p} - F_{T,1}^{q/p} \big) \hat \C_{9,q}^{(7)} \,,
\\
\label{eq:cNR4}
c_{4}^p &= -\frac{2\alpha}{\pi}\frac{\mu_p}{m_N} \hat\C_1^{(5)}+ \sum_q \Big( 8 F_{T,0}^{q/p} \, \hat
\C_{9,q}^{(7)} - 4 F_{A}^{q/p} \, \hat \C_{4,q}^{(6)} \Big) \,,
\\
\label{eq:cNR5}
c_{5}^p &=\frac{2\alpha Q_p m_N}{\pi \vec q^{\,\,2}}\hat \C_q^{(5)},
\\
\label{eq:cNR6}
c_{6}^p &= \frac{2\alpha}{\pi\vec q^{\,\,2}}\mu_p m_N\hat \C_a^{(5)}+\sum_q \Big( F_{P'}^{q/p} \, \hat
\C_{4,q}^{(6)} + \frac{m_N}{m_\chi} F_{P}^{q/p} \, \hat \C_{8,q}^{(7)}
\Big) + \frac{m_N}{m_\chi} F_{\tilde G}^{p} \, \hat \C_{4}^{(7)} \,,
\\
\label{eq:cNR7}
c_{7}^p &= - 2 \sum_q F_{A}^{q/p} \, \hat \C_{3,q}^{(6)} \,,
\\
\label{eq:cNR8}
c_{8}^p &= 2 \sum_q F_{1}^{q/p} \, \hat \C_{2,q}^{(6)} \,,
\\
\label{eq:cNR9}
c_{9}^p &= 2 \sum_q \Big[ \big( F_{1}^{q/p} + F_{2}^{q/p}
\big) \, \hat \C_{2,q}^{(6)} + \frac{m_N}{m_\chi} F_{A}^{q/p} \, \hat
\C_{3,q}^{(7)} \Big] \,,
\\
\label{eq:cNR10}
c_{10}^p &= F_{\tilde G}^p \, \hat \C_{3}^{(7)} + \sum_q
\Big( F_{P}^{q/p} \, \hat \C_{7,q}^{(7)} - 2 \frac{m_N}{m_\chi}
F_{T,0}^{q/p} \, \hat \C_{10,q}^{(7)} \Big) \,,
\\
\label{eq:cNR11}
c_{11}^p &= \frac{2\alpha}{\pi}Q_p \frac{m_N}{\vec q^{\,\,2}}\hat \C_2^{(5)}+ \sum_q \Big[ 2 \big( F_{T,0}^{q/p} -
  F_{T,1}^{q/p} \big) \, \hat \C_{10,q}^{(7)} - \frac{m_N}{m_\chi}
  F_{S}^{q/p} \, \hat \C_{6,q}^{(7)} \Big] - \frac{m_N}{m_\chi}
F_{G}^p \, \hat \C_{2}^{(7)} \,, 
\\ 
c_{12}^p&= - 8 \sum_q
F_{T,0}^{q/p} \, \hat \C_{10,q}^{(7)} \,.
\end{align}
Here $Q_{p(n)}=1(0)$ is the charge of the proton (neutron), while
$\alpha$ is the electromagnetic fine structure constant. The sums run
over the light quark flavors $q=u,d,s$. The coefficients for neutrons
are obtained by replacing $p\to n$, $u\leftrightarrow d$. In the above
expressions the nonperturbative effects of the strong interactions is
encoded in the form factors for the single-nucleon currents,
\begin{align}
\label{vec:form:factor}
\langle N'|\bar q \gamma^\mu q|N\rangle&=\bar u_N'\Big[F_1^{q/N}(q^2)\gamma^\mu+\frac{i}{2m_N}F_2^{q/N}(q^2) \sigma^{\mu\nu}q_\nu\Big]u_N\,,
\\
\label{axial:form:factor}
\langle N'|\bar q \gamma^\mu \gamma_5 q|N\rangle&=\bar u_N'\Big[F_A^{q/N}(q^2)\gamma^\mu\gamma_5+\frac{1}{2m_N}F_{P'}^{q/N}(q^2) \gamma_5 q^\mu\Big]u_N\,,
\\
\label{scalar:form:factor}
\langle N'| m_q \bar q   q|N\rangle&= F_S^{q/N} (q^2)\, \bar u_N' u_N\,,
\\
\label{pseudoscalar:form:factor}
\langle N'| m_q \bar q  i \gamma_5 q|N\rangle&= F_P^{q/N} (q^2)\, \bar u_N' i \gamma_5 u_N\,,
\\
\label{CPeven:gluonic:form:factor}
\langle N'| \frac{\alpha_s}{12\pi} G^{a\mu\nu}G^a_{\mu\nu} |N\rangle&= F_G^{N} (q^2)\, \bar u_N' u_N\,,
\\
\label{CPodd:gluonic:form:factor}
\langle N'| \frac{\alpha_s}{8\pi} G^{a\mu\nu}\tilde G^a_{\mu\nu}|N\rangle&= F_{\tilde G}^{N} (q^2)\, \bar u_N' i \gamma_5 u_N\,,
\\
\label{tensor:form:factor}
\begin{split}
\langle N'|m_q \bar q \sigma^{\mu\nu} q |N\rangle&=  \bar u_N'\Big[F_{T,0}^{q/N} (q^2)\,  \sigma^{\mu\nu} +\frac{i}{2 m_N} \gamma^{[\mu}q^{\nu]} F_{T,1}^{q/N} (q^2) 
\\
&\qquad \qquad+ \frac{i}{m_N^2} q^{[\mu}k_{12}^{\nu]} F_{T,2}^{q/N} (q^2) \Big] u_N\,.
\end{split}
\end{align}
Here we shortened $\langle N'|=\langle N(k_2)| $, $| N\rangle= |
N(k_1)\rangle $, $\bar u_N'= u_N(k_2)$, $u_N= u_N(k_1)$ and introduced
$q^\mu=k_2^\mu-k_1^\mu$, $k_{12}^\mu=k_1^\mu+k_2^\mu$. Expanding the
form factors to LO in chiral counting, the expressions for the axial
current, the pseudoscalar current, and the CP-odd gluonic current
contain light-meson poles,
\begin{align}
\label{eq:F_PP'}
F_{i}^{q/N}(q^2)&=\frac{m_N^2}{m_\pi^2-q^2} a_{i,\pi}^{q/N}+\frac{m_N^2}{m_\eta^2-q^2} a_{i,\eta}^{q/N}
+\cdots\,, \quad i=P,P'\,,
\\
\label{eq:F_tildeG}
F_{\tilde G}^{N}(q^2)&=
\frac{q^2}{m_\pi^2-q^2} a_{\tilde G,\pi}^{N}+\frac{q^2}{m_\eta^2-q^2} a_{\tilde G,\eta}^{N}+b_{\tilde G}^{N}
+\cdots\,,
\end{align}
while all the other form factors can be evaluated at $q^2=0$,
\begin{equation}
\label{eq:Fi}
F_{i}^{q/N}(q^2)=F_{i}^{q/N}(0)+\cdots\,.
\end{equation}
The ellipses denote terms of higher order in chiral counting. Below we
collect the expressions for the proton form factors with further
details given in Ref. \cite{Bishara:2017pfq}. We work in the isospin
limit, so that $F_i^{u(d,s)/p}= F_i^{d(u,s)/n}$, with the exception of
the scalar form factors, where we give the values separately for
proton and neutron, and of the tensor form factors, where the isospin
relations involve quark masses (see below). The numerical input for
the hadronic parameters is collected in Tab.~\ref{tab:inputs}.  In the
\ddm~package, these parameters are set in the file
\ddm{\ttfamily/inputs.m} in case the user wishes to update the values
or set different ones.

\begin{table}
\begin{center}
\begin{tabular}{cccccc}
\hline\hline
parameter & value & parameter & value & parameter & value
 \\
\hline
$\mu_p$&$2.793$ 		&$\sigma_u^p$&$(17\pm 5)$~MeV		& $g_T^u$	& $0.794(15)$
\\
$\mu_n$&$-1.913$ 		&$\sigma_d^p$&$(32\pm 10)$~MeV		& $g_T^d$	& $-0.204(8)$
\\ 
$g_A$&$1.2723(23)$	&$\sigma_u^n$&$(15\pm5)$~MeV		& $g_T^s$	& $(3.2\pm8.6)\cdot10^{-4}$
\\ 
$\Delta u+\Delta d$&$0.521(53)\,$ &$\sigma_d^n$&$(36\pm10)$~MeV& $B_{T,10}^{u/p}$ & $3.0\pm1.5$
\\
$\Delta s$&$-0.031(5)$	&$\sigma_s^{p,n}$&$(41.3\pm7.7)$~MeV	& $B_{T,10}^{d/p}$ & $0.24\pm0.12$
\\ 
$m_u/m_d$ & $0.46(5)$	&$m_G$		&$848(14)~{\rm MeV}$		& $B_{T,10}^{s/p}$ & $0.0\pm0.2$
\\
$2 m_s/(m_u+m_d)$ & $27.5(3)$	& & & &
\\
\hline
\hline
\end{tabular}
\end{center}
\caption{ \label{tab:inputs} The hadronic parameters used in
  evaluating the form factors in the code (see main text for details).
}
\end{table}

{\em Vector current.} The Dirac form factors at zero recoil count the
number of valence quarks in the nucleon, thus $F_1^{u/p}(0)=2$,
$F_1^{d/p}(0)=1$, and $F_1^{s/p}(0)=0$. The Pauli form factors for $u$
and $d$ quarks are $F_2^{u/p}(0)=2 (\mu_p-1)+\mu_n+F_2^{s/p}(0)$,
$F_2^{d/p}(0)=2 \mu_n+(\mu_p-1)+F_2^{s/p}(0)$, where we use as inputs
the proton and neutron magnetic moments, $\mu_p\simeq 2.793$,
$\mu_n\simeq-1.913$, and the Pauli form factor for the $s$ quark,
$F_2^{s/p}(0)= -0.064(17)$~\cite{Sufian:2016pex}.

{\em Axial current.} The axial form factor at zero recoil is
$F_A^{q/p}(0)=\Delta q$. As numerical inputs we use $g_A=\Delta
u-\Delta d=1.2723(23)$~\cite{Olive:2016xmw}, and, in the
$\overline{\rm MS}$ scheme at $Q=2$~GeV, $\Delta u+\Delta
d=0.521(53)$~\cite{diCortona:2015ldu}, $\Delta s=-0.031(5)$
\cite{QCDSF:2011aa, Engelhardt:2012gd, Bhattacharya:2015gma,
  Alexandrou:2017hac}.  The residua of the pion- and eta-pole
contributions to $F_{P'}^{q/N}$ are $a_{P',\pi}^{u/p} =
-a_{P',\pi}^{d/p} = 2 g_A$\,, $a_{P',\pi}^{s/p} = 0$, and
$a_{P',\eta}^{u/p} = a_{P',\eta}^{d/p} = -a_{P',\eta}^{s/p}/2 =
2\big(\Delta u+\Delta d-2 \Delta s\big)/3$, respectively.

{\em Scalar current.}  The scalar form factors at zero recoil are
conventionally referred to as nuclear sigma terms, $F_S^{q/N}(0)=
\sigma_q^N$. We use $\sigma_u^p=(17\pm 5){\rm~MeV}$,
$\sigma_d^p=(32\pm 10){\rm~MeV}$, $\sigma_u^n=(15\pm 5){\rm~MeV}$\,,
$\sigma_d^n=(36\pm 10){\rm~MeV}$, obtained from expressions in
Ref.~\cite{Crivellin:2013ipa} using a rather conservative estimate
$\sigma_{\pi N}=(50\pm15)$ MeV \cite{Bishara:2017pfq}, along with $
\sigma_s^p=\sigma_s^n=(41.3\pm 7.7){\rm~MeV}$~\cite{Junnarkar:2013ac,
  Yang:2015uis, Durr:2015dna}.

{\em Pseudoscalar current.}  The residua of the light-meson poles for
the pseudoscalar form factors, $F_{P}^{q/N}(q^2)$, are given by $
a_{P,\pi}^{u/p}/{m_u} = -a_{P,\pi}^{d/p}/{m_d} = g_A B_0/m_N$,
$a_{P,\pi}^{s/p} = 0$, $a_{P,\eta}^{u/p}/{m_u} =
a_{P,\eta}^{d/p}/{m_d}=-a_{P,\eta}^{s/p}/(2{m_s}) = B_0\big(\Delta
u_p+\Delta d_p-2 \Delta u_s\big)/(3m_N)$, where $B_0$ is a ChPT
constant, related to the quark condensate, that always appears
multiplied by a quark mass, $B_0 m_q$. In the code we re-express these
as $B_0 m_u = m_\pi^2/(1+m_d/m_u)= (6.1\pm0.5) \times 10^{-3}
\,\text{GeV}^2$, $B_0 m_d = m_\pi^2/(1+m_u/m_d) = (13.3\pm0.5) \times
10^{-3} \,\text{GeV}^2$, and $B_0 m_s =m_\pi^2 m_s/(m_u+m_d) = (268
\pm 3) \times 10^{-3} \,\text{GeV}^2$, using $m_u/m_d=0.46\pm0.05$ and
$2m_s/(m_u+m_d) = 27.5\pm0.3$~\cite{Olive:2016xmw}.

{\em CP-even gluonic current.} The value of the relevant form factor
at zero recoil is given by $F_G^{N}(0)=- 2m_G/27$, where
$m_G=m_N-\sum_q \sigma_q^N=(848\pm14) {\rm ~MeV}$, using the values of
nuclear sigma terms in Table \ref{tab:inputs}.

{\em CP-odd gluonic current.} The parameters describing the CP-odd
gluonic form factor in \eqref{eq:F_tildeG} can be expressed in terms
of the matrix elements of the axial current, and are given by $2
a_{\tilde G,\pi}^{N}=-\tilde m m_N g_A \big(1/m_u-1/m_d\big)$,
$6a_{\tilde G,\eta}^{N}=-\tilde m m_N \big(\Delta u+\Delta d-2 \Delta
s)\big(1/m_u+1/m_d-2/m_s\big)$, $b_{\tilde G}^{N}=-\tilde m m_N
\big(\sum_q \Delta q/m_q\big)$, where $1/\tilde
m=(1/m_u+1/m_d+1/m_s)$.

{\em Tensor current.} The matrix elements of tensor currents are
described by three sets of form factors, but only two enter the
chirally leading expressions, $F_{T,0}^{q/p}(0)=m_q g_{T}^{q}$, and $
F_{T,1}^{q/N}(0)= - m_q B_{T,10}^{q/N} (0)$.  In the $\overline{\rm
  MS}$ scheme at $Q=2$~GeV, one has $g_T^u=0.794\pm0.015$,
$g_T^d=-0.204\pm0.008$, $g_T^s=(3.2\pm8.6)\cdot 10^{-4}$
\cite{Alexandrou:2017qyt, Bhattacharya:2016zcn}. Using the results of
the constituent quark model in~\cite{Pasquini:2005dk} we estimate
$B_{T,10}^{u/p}(0)=3.0\pm1.5$, $B_{T,10}^{d/p}(0)= 0.24\pm0.12$, and
$B_{T,10}^{s/p}=0.0\pm0.2$. For neutrons one has
$F_{T,0}^{u(d,s)/n}(0)=m_{u(d,s)} g_{T}^{d(u,s)}$ and $
B_{T,10}^{u(d,s)/n} (0)=B_{T,10}^{d(u,s)/p} (0)$.

\subsection{QCD and QED running}
\label{sec:RG}
The Wilson coefficients for the \nfiv~and \nfv~bases are defined at
$\mu_b=m_b$ and $\mu_Z=M_Z$, respectively, and need to be evolved down
to $\mu_h=2\,$GeV, where the matching to the hadronic theory is
performed. The RG evolution is achieved by standard methods (see,
e.g., Ref.~\cite{Hill:2014yxa}) and involves the running from $\mu_Z$
to $\mu_b$ in the five-flavor scheme, and from $\mu_b$ to $\mu_h$ in
the four-flavor scheme, integrating out the $b$ quark and the $c$
quark at the two thresholds.

Since the DM fields are QCD and QED singlets, the RG evolution of the
operators
Eqs.~\eqref{eq:dim5:nf5:Q1Q2:light}-\eqref{eq:dim7:Q9Q10:light} is due
only to their SM fields. Several of the operators have vanishing
anomalous dimensions and the associated Wilson coefficients are RG
invariant: this is the case for the dipole operators
Eq.~\eqref{eq:dim5:nf5:Q1Q2:light}, the operators involving a quark
vector current Eq.~\eqref{eq:dim6:Q1Q2:light}, and the scalar
operators Eq.~\eqref{eq:dim7:Q5Q6:light}. Moreover, there is no
one-loop QCD running for the operators involving an axial-vector quark
current Eq.~\eqref{eq:dim6:Q3Q4:light}, the pseudoscalar operators
Eq.~\eqref{eq:dim7:Q7Q8:light}, and the gluonic operators
Eqs.~\eqref{eq:dim7:Q1Q2:light}-\eqref{eq:dim7:Q3Q4:light} , so the
only relevant effect of the RG evolution is a (small) rescaling of the
coefficients of the tensor operators Eq.~\eqref{eq:dim7:Q9Q10:light},
and the mixing of the gluonic operators
Eqs.~\eqref{eq:dim7:Q1Q2:light}-\eqref{eq:dim7:Q3Q4:light} into the
scalar operators
Eqs.~\eqref{eq:dim7:Q5Q6:light}-\eqref{eq:dim7:Q7Q8:light}.

\begin{figure}[t]\centering
	\includegraphics[scale=0.8]{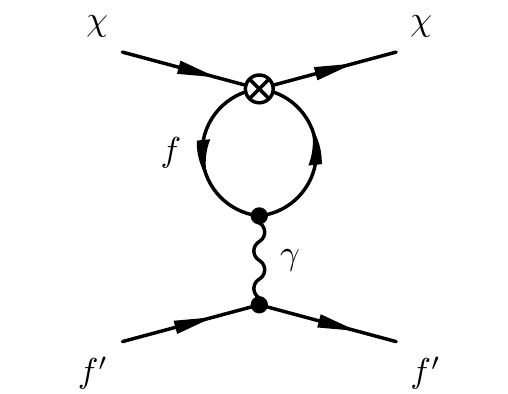}
	\caption{Mixing of dimension-six four-fermion operators into
          each other via the photon penguin insertion.}
	\label{fig:d6-photon-peng}
\end{figure}

The QED contributions to the RG evolution can, in general, be
neglected, due to the smallness of the electromagnetic coupling
constant. The only exception is the off-diagonal mixing induced by
photonic penguin diagrams (see Fig.~\ref{fig:d6-photon-peng}) of the
operators $\Q_{1,f}^{(6)}$ or $\Q_{2,f}^{(6)}$ among themselves, for
different fermion flavors $f$. In this way, scattering on atomic
nuclei can be generated even if, at tree level, DM couples only to
leptons~\cite{Kopp:2009et}.  Note that the conservation of parity
forbids the mixing of $\Q_{1,f}^{(6)}$ into $\Q_{2,f}^{(6)}$, or vice
versa. The penguin insertions for all operators other
than Eq.~\eqref{eq:dim6:Q1Q2:light} vanish.

Finite corrections arise at each heavy flavor threshold. Beside the
usual threshold corrections to $\alpha_s$ (see, e.g.,
Ref.~\cite{Chetyrkin:1997un}), there are also finite threshold
corrections for the operators
Eq.~\eqref{eq:dim7:Q1Q2:light}-\eqref{eq:dim7:Q3Q4:light}, where at
$\mu=\mu_b$,
\begin{equation}
\begin{split}\label{eq:threshold:mb}
  \hat \C_{1(2)}^{(7)}|_{n_f=4} (\mu_b) &= \hat \C_{1(2)}^{(7)}|_{n_f=5} (\mu_b) - 
  \hat \C_{5,b(6,b)}^{(7)}|_{n_f=5} (\mu_b) \,, \\
  \hat \C_{3(4)}^{(7)}|_{n_f=4} (\mu_b) &= \hat \C_{3(4)}^{(7)}|_{n_f=5} (\mu_b) +
  \hat \C_{7,b(8,b)}^{(7)}|_{n_f=5} (\mu_b) \,, 
\end{split}
\end{equation}
while at $\mu=\mu_c$,
\begin{equation}
\begin{split}\label{eq:threshold:mc}
  \hat \C_{1(2)}^{(7)}|_{n_f=3} (\mu_c) &= \hat \C_{1(2)}^{(7)}|_{n_f=4} (\mu_c) - 
  \hat \C_{5,c(6,c)}^{(7)}|_{n_f=4} (\mu_c) \,, \\
  \hat \C_{3(4)}^{(7)}|_{n_f=3} (\mu_c) &= \hat \C_{3(4)}^{(7)}|_{n_f=4} (\mu_c) +
  \hat \C_{7,c(8,c)}^{(7)}|_{n_f=4} (\mu_c) \,,
\end{split}
\end{equation}
such that the effects of the heavy quarks appear, at low energies, as
additional contributions to the gluonic
operators Eq.~\eqref{eq:dim7:Q1Q2:light}-\eqref{eq:dim7:Q3Q4:light}. All
other Wilson coefficients cross the thresholds continuously, $\hat
\C_i^{(d)}|_{n_f-1} = \hat \C_i^{(d)}|_{n_f}$.

\subsection{Scalar dark matter}
For scalar DM, the effective interactions with the SM start at
dimension six,
\begin{equation}\label{eq:lightDM:Ln:scalar}
{\cal L}_\varphi=
\hat \C_{a}^{(6)} {\cal Q}_a^{(6)}+\cdots\,, 
\qquad {\rm where}\quad 
\hat \C_{a}^{(6)}=\frac{\C_{a}^{(6)}}{\Lambda^{2}}\,.
\end{equation}
Again, the $\C_{a}^{(6)}$ here are the dimensionless Wilson
coefficients\footnote{For operators and Wilson coefficients we adopt
  the same notation for scalar DM as for fermionic DM. No confusion
  should arise as this abuse of notation is restricted to this
  section.  In our code, the user can select either fermionic or
  scalar DM, see Sec.~\ref{sec:program}.} of the effective
interactions between DM and the SM. The operators coupling DM to
quarks and gluons are
\begin{align}
\label{eq:dim6:Q1Q2:light:scalar}
{\cal Q}_{1,f}^{(6)} & = \big(\varphi^* i\overset{\leftrightarrow}{\partial_\mu} \varphi\big) (\bar f \gamma^\mu f)\,,
 &{\cal Q}_{2,f}^{(6)} &= \big(\varphi^* i\overset{\leftrightarrow}{\partial_\mu} \varphi\big)(\bar f \gamma^\mu \gamma_5 f)\,, 
 \\
 \label{eq:dim6:Q3Q4:light:scalar}
 {\cal Q}_{3,f}^{(6)} & = m_f (\varphi^* \varphi)( \bar f f)\,, 
&{\cal
  Q}_{4,f}^{(6)} &= m_f (\varphi^* \varphi)( \bar f i\gamma_5 f)\,,
\\
\label{eq:dim6:Q5Q6:light:scalar}
  {\cal Q}_5^{(6)} & = \frac{\alpha_s}{12\pi} (\varphi^* \varphi)
 G^{a\mu\nu}G_{\mu\nu}^a\,, 
 & {\cal Q}_6^{(6)} &= \frac{\alpha_s}{8\pi} (\varphi^* \varphi) G^{a\mu\nu}\widetilde G_{\mu\nu}^a\,,
\end{align}
while the couplings to photons are\footnote{Note that the operator
  with one electromagnetic field strength tensor, $\partial_\mu
  \varphi^* \partial_\nu\varphi F^{\mu\nu}$, can be reduced to
  $\Q_{1,f}^{(6)}$ by using equations of motion for the photon field.}
\begin{align}
  \label{eq:dim6:Q7Q8:light:scalar} 
 {\cal Q}_{7}^{(6)} &=\frac{\alpha}{\pi}(\varphi^*
\varphi) F^{\mu\nu}F_{\mu\nu}\,, 
& {\cal Q}_{8}^{(6)} &=\frac{3\alpha}{\pi}(\varphi^*
\varphi) F^{\mu\nu} \tilde F_{\mu\nu}\,. &~&
\end{align}
Here $\overset{\leftrightarrow}{\partial_\mu}$ is defined through
$\phi_1\overset{\leftrightarrow}{\partial_\mu} \phi_2=\phi_1
\partial_\mu \phi_2- (\partial_\mu \phi_1) \phi_2$. The operators
${\cal Q}_4^{(6)}$, ${\cal Q}_6^{(6)}$, and ${\cal Q}_8^{(6)}$ are
CP-odd, while all the other operators are CP-even. In complete analogy
to the case of fermionic DM, the operator basis can be specified in
the \nfiii, \nfiv, or \nfv{} scheme. The running of the Wilson
coefficients and matchings at heavy flavor thresholds proceed along
the lines discussed in Sec.~\ref{sec:RG}.

The \ddm{} code provides the matching onto the nonrelativistic
Lagrangian for interactions with nucleons, ${\cal L}_{\rm
  NR}=\sum_{i,N} c_i^N(q^2) \op_i^N$, also for scalar DM. For the
basis of nonrelativistic operators we use the same basis as for
fermionic DM, Eqs.~\eqref{eq:O1pO4p}-\eqref{eq:O11pO12p}, dropping all
operators that involve the DM spin. The coefficients $c_i^N(q^2)$
are~\cite{Bishara:2017pfq}
\begin{align}
c_1^N&=\sum_q\Big(2 m_\varphi F_1^{q/N}  \hat \C_{1q}^{(6)}+F_S^{q/N} \hat \C_{3q}^{(6)}\Big) +F_G \hat \C_{5}^{(6)}\,,
\\
c_7^N&= - 4 m_\varphi \sum_q F_A^{q/N}  \hat \C_{2q}^{(6)}\,,
\\
c_{10}^N&= \sum_q F_P^{q/N} \hat \C_{4q}^{(6)}+ F_{\tilde G}  \hat \C_{6}^{(6)}\,,
\end{align}
where the sum again runs over the light quark fields, $q=u,d,s$. The
nuclear matrix elements of the electromagnetic operators
\eqref{eq:dim6:Q7Q8:light:scalar} are rather uncertain and are
currently not implemented in \ddm.

\section{The program}
\label{sec:program}
The matching and running described above are implemented in a
\mma~package available at
\url{https://directdm.github.io}.
The \ddm~package has been tested on \mma{} versions 10 and 11.
The package can be loaded via:
\begin{Verbatim}[frame=single,xleftmargin=1cm,xrightmargin=1cm,commandchars=\\\{\}]
  $DirectDMDirectory="</path/to/directdm/directory>";
  <<DirectDM`
\end{Verbatim}

By default, the DM is assumed to be a Dirac fermion. This setting can
be changed with the function
\begin{Verbatim}[frame=single,xleftmargin=1cm,xrightmargin=1cm,commandchars=\\\{\}]
  SetDMType["type"]
\end{Verbatim}
where \verb|"type"| can be \verb|"D"| for a Dirac fermion, \verb|"M"|
for a Majorana fermion, \verb|"C"| for a complex scalar, and
\verb|"S"| for a real scalar.

Once loaded, the user can set the Wilson coefficients in the desired
initial basis. The package then performs the running in the
intermediate EFTs and the matching at the intermediate thresholds
until the user-specified final basis is reached. The available bases
are: \nfiii, \nfiv, \nfv, and \nr. The syntax to set the Wilson
coefficients in the $n_f=\{3,4,5\}$ bases is:
\begin{Verbatim}[frame=single,xleftmargin=1cm,xrightmargin=1cm,commandchars=\\\{\}]
  SetCoeff["basis",QD[i,f],value]
\end{Verbatim}
\noindent The allowed arguments are \verb|basis| $\in$ \{\nfiii,
\nfiv, \nfv\}, \verb|QD| $\in$ \{{\ttfamily Q5, Q6, Q7}\} is the mass
dimension of the operator, \verb|i| is the operator number, and,
finally, \verb|f| $\in$ \{{\ttfamily "u", "d", "s", "c", "b", "e",
  "mu", "tau"}\} is the flavor index of the operator where the allowed
values clearly depend on the basis in question. These operators are
defined in
Eqs.~(\ref{eq:dim5:nf5:Q1Q2:light})-(\ref{eq:dim7:Q9Q10:light}). Note
that in the case that the operator does not include a SM fermion
current, the operator name syntax is simply \verb|QD[i]| with no
flavor index.
\begin{table}[t]\centering
	\renewcommand{\arraystretch}{1.15}
	\setlength{\tabcolsep}{1em}
	\begin{tabular}{c c c c}\hline\hline
		DM Type & \verb|SetDMType| & Operator Numbering & Basis Definition\\\hline
		Dirac fermion & \verb|"D"| & 
		$\begin{matrix}
		\texttt{Q5}\,, & i\in\{1,2\}\\
		\texttt{Q6}\,, & i\in\{1,\dots,4\}\\
		\texttt{Q7}\,, & i\in\{1,\dots,10\}
		\end{matrix}$ & Eqs.~(\ref{eq:dim5:nf5:Q1Q2:light})-(\ref{eq:dim7:Q9Q10:light})\\\hline
		Majorana fermion & \verb|"M"| &
		$\begin{matrix}
		\texttt{Q6}\,, & i\in\{2,4\}\\
		\texttt{Q7}\,, & i\in\{1,\dots,8\}
		\end{matrix}$&
		Eqs.~(\ref{eq:dim6:Q2Q4:majorana})-(\ref{eq:dim7:Q7Q8:majorana})\\\hline
		Complex scalar & \verb|"C"| &
		$\begin{matrix}\texttt{Q6}\,, & i\in\{1,\dots,6\}\end{matrix}$&
		Eqs.~(\ref{eq:dim6:Q1Q2:light:scalar})-(\ref{eq:dim6:Q5Q6:light:scalar})\\\hline
		Real scalar & \verb|"R"| &
		$\begin{matrix}\texttt{Q6}\,, & i\in\{3,\dots\,6\}\end{matrix}$ &
		Eqs.~(\ref{eq:dim6:Q3Q4:scalar:real})-(\ref{eq:dim6:Q7Q8:scalar:real})\\
		\hline\hline
	\end{tabular}
	\caption{Operator dimensions and numbering for the Dirac/Majorana fermion and complex/real scalar bases. Setting operator numbers outside the allowed values shown here will cause the {\ttfamily SetCoeff} function to generate an error message.}
	\label{tab:operator-numbering}
\end{table}
The allowed values for the indices \verb|i| depend on the type of DM and the operator dimension. They are given in Tab.~\ref{tab:operator-numbering}.
The matching scales are:
\begin{equation}
\begin{split}
\texttt{5Flavor}:\qquad \mu_Z&= M_Z = 91.1876\,\text{GeV}\,,\\
\texttt{4Flavor}:\qquad  \mu_b&= 4.18\,\text{GeV}\,,\\
\texttt{3Flavor}:\qquad \mu_c&= 2\,\text{GeV}\,,\\
\texttt{NR}:\qquad \mu_N&= 2\,\text{GeV}\,.
\end{split}
\end{equation}
Consequently there is no running in the \nfiii~flavor basis.

To perform the running and matching, the user must then issue the command
\begin{Verbatim}[frame=single,xleftmargin=1cm,xrightmargin=1cm,commandchars=\\\{\}]
  ComputeCoeffs["basis_i", "basis_f"]
\end{Verbatim}
where \verb|basis_i| is the intial basis and \verb|basis_f| is the
final basis. The \verb|ComputeCoeffs| function takes an optional
argument \verb|Running -> True/False|. It is set to \verb|True| by
default. Setting it to \verb|False| disables the QCD and QED running
in the intermediate EFTs. As mentioned above, this option has no
effect in the \nfiii~basis.

Finally, to retrieve the Wilson coefficients in the final basis, the
package provides two functions: \verb|GetCoeff| and
\verb|CoeffsList|. The former takes the same arguments as the
\verb|SetCoeff| function but is only implemented for the \nfv, \nfiv,
and \nfiii~bases.
\begin{Verbatim}[frame=single,xleftmargin=1cm,xrightmargin=1cm,commandchars=\\\{\}]
  GetCoeff["basis",QD[i,f]]
\end{Verbatim}
This function allows the user to retrieve one coefficient at a
time. For the {\nr} basis, however, it is more practical to retrieve
the entire list of Wilson coefficients. This can be done with:
\begin{Verbatim}[frame=single,xleftmargin=1cm,xrightmargin=1cm,commandchars=\\\{\}]
  CoeffsList["basis"]
\end{Verbatim}
where \verb|basis|, in this case, can be \nfv, \nfiv, \nfiii,
\verb|NR_p|, or \verb|NR_n|. Note the syntax for the \verb|NR| basis;
here, the user must specify the proton, \verb|"NR_p"|, or the neutron,
\verb|"NR_n"|, basis explicitly.

Of course, the user might wish to set all Wilson coefficients to zero
and start afresh. This can be done via:
\begin{Verbatim}[frame=single,xleftmargin=1cm,xrightmargin=1cm,commandchars=\\\{\}]
  ResetBasis["basis"]
\end{Verbatim}
where \verb|basis| can take any of the allowed values discussed
above. For the NR EFT, \verb|basis| can only be \verb|NR| -- i.e., not
\verb|NR_p| or \verb|NR_n|.  If called without an argument, i.e.,
\verb|ResetBasis[]|, the function resets all bases.

Output of the {\tt DirectDM} code is structured in such a way that it
is easy to interface with the {\tt DMFormFactor} package
\cite{Anand:2013yka}. An example of such an interface is given in the
     {\tt example.nb} notebook, included in the distribution.

\section{Conclusions}
\label{sec:conclusion}
We presented a {\tt Mathematica} package, {\tt DirectDM}, that
performs an important intermediate step in the calculation of event
rates in dark matter direct detection experiments. It takes as an
input the Wilson coefficients of the EFT coupling DM to quark, gluons,
photons, and performs a matching onto an EFT describing DM with
nonrelativistic protons and neutrons, at leading order in a chiral
expansion of the hadronic form factors. The QCD and QED RG evolution
from the electroweak to the hadronic scale, finite matching
corrections at the heavy quark thresholds, and tree-level meson
exchange contributions in the chiral effective theory are consistently
taken into account. The effects of operator mixing above the
electroweak scale will be included as part of a future
project~\cite{future:BBGZ}.

{\bf Acknowledgements.} FB is supported by the Science and Technology
Facilities Council (STFC). JZ is supported in part by the
U.S. National Science Foundation under CAREER Grant PHY- 1151392 and
by the DOE grant de-sc0011784. BG is supported in part by the
U.S. Department of Energy under grant DE-SC0009919.

\appendix
\section{Operator basis for Majorana and real scalar DM}
\label{app:majorana-and-real-scalar}

\paragraph{\it Majorana fermion.} For simplicity we use the same
notation for the operators with Majorana fermion DM and for the
operators with Dirac fermion DM, if the Lorentz structures of the
DM~$\otimes$~SM currents coincide. For a Majorana fermion the
operators $\Q_{1,2}^{(5)}$, $\Q_{1,f}^{(6)}$, $\Q_{3,f}^{(6)}$,
$\Q_{9,f}^{(7)}$, and $\Q_{10,f}^{(7)}$ are absent since in that case
the vector and tensor currents vanish. We include an additional factor
of 1/2 in the definition of the Majorana DM operators to compensate
for the additional Wick contraction in the case of a Majorana fermion.
The two nonzero dimension six operators are,
\begin{align}
{\cal Q}_{2,f}^{(6)} &= \frac{1}{2}\,(\bar \chi\gamma_\mu\gamma_5 \chi)(\bar f \gamma^\mu f)\,,
& {\cal Q}_{4,f}^{(6)}& = \frac{1}{2}\,(\bar
\chi\gamma_\mu\gamma_5 \chi)(\bar f \gamma^\mu \gamma_5 f)\,,
\label{eq:dim6:Q2Q4:majorana}
\end{align}
and the eight nonzero dimension seven operators, 
\begin{align}
{\cal Q}_1^{(7)} & = \frac{\alpha_s}{24\pi} (\bar \chi \chi)
G^{a\mu\nu}G_{\mu\nu}^a\,, 
& {\cal Q}_2^{(7)} &= \frac{\alpha_s}{24\pi} (\bar \chi i\gamma_5 \chi) G^{a\mu\nu}G_{\mu\nu}^a\,,\label{eq:dim7:Q1Q2:majorana}
\\
{\cal Q}_3^{(7)} & = \frac{\alpha_s}{16\pi} (\bar \chi \chi) G^{a\mu\nu}\widetilde
G_{\mu\nu}^a\,, 
& {\cal Q}_4^{(7)}& = \frac{\alpha_s}{16\pi}
(\bar \chi i \gamma_5 \chi) G^{a\mu\nu}\widetilde G_{\mu\nu}^a \,, \label{eq:dim7:Q3Q4:majorana}
\\
{\cal Q}_{5,f}^{(7)} & =\frac{ m_f}{2} (\bar \chi \chi)( \bar f f)\,, 
&{\cal
	Q}_{6,f}^{(7)} &= \frac{ m_f}{2} (\bar \chi i \gamma_5 \chi)( \bar f f)\,,\label{eq:dim7:Q5Q6:majorana}
\\
{\cal Q}_{7,f}^{(7)} & = \frac{ m_f}{2}\,(\bar \chi \chi) (\bar f i \gamma_5 f)\,, 
&{\cal Q}_{8,f}^{(7)} & = \frac{ m_f}{2} (\bar \chi i \gamma_5 \chi)(\bar f i \gamma_5
f)\,, \label{eq:dim7:Q7Q8:majorana}  
\end{align}

\paragraph{Real scalar.} Similarly, the operators for a real scalar DM are a subset
of the operators for complex scalar DM, and carry an additional factor of 1/2.
The relevant dimension six operators are,
\begin{align}
\label{eq:dim6:Q3Q4:scalar:real}
{\cal Q}_{3,f}^{(6)} & = \frac{m_f}{2} (\varphi\varphi)( \bar f f)\,, 
&{\cal
	Q}_{4,f}^{(6)} &= \frac{m_f}{2} (\varphi\varphi)( \bar f i\gamma_5 f)\,,
\\
\label{eq:dim6:Q5Q6:scalar:real}
{\cal Q}_5^{(6)} & = \frac{\alpha_s}{24\pi} (\varphi\varphi)
G^{a\mu\nu}G_{\mu\nu}^a\,, 
& {\cal Q}_6^{(6)} &= \frac{\alpha_s}{16\pi} (\varphi\varphi) G^{a\mu\nu}\widetilde G_{\mu\nu}^a\,,\\
\label{eq:dim6:Q7Q8:scalar:real} 
{\cal Q}_{7}^{(6)} &=\frac{\alpha}{2\pi}(\varphi
\varphi) F^{\mu\nu}F_{\mu\nu}\,, 
& {\cal Q}_{8}^{(6)} &=\frac{3\alpha}{4\pi}(\varphi
\varphi) F^{\mu\nu} \tilde F_{\mu\nu}\,. &~&
\end{align}

\section{Translation from the basis of Goodman et al.}\label{sec:tait}

Here we provide a translation between our basis for DM interactions,
Eqs.~\eqref{eq:dim5:nf5:Q1Q2:light}-\eqref{eq:dim7:Q9Q10:light} and
Eqs.~\eqref{eq:dim6:Q1Q2:light:scalar}-\eqref{eq:dim6:Q7Q8:light:scalar},
to the basis used by Goodman et al., Ref.~\cite{Goodman:2010qn}.
For {\em Dirac fermion DM} the EFT
interaction Lagrangian in the basis of Ref.~\cite{Goodman:2010qn} is
\begin{equation}
\label{eq:Goodman:Dirac}
{\cal L}_\chi=\sum_{i} \, G_{Di} \Q_{Di},
\end{equation}
where the operators $\Q_{Di}$, $i=1,\ldots,16$ (the Wilson
coefficients $G_{Di}$) are listed in the 2nd (3rd) column of Table~II,
left, in Ref.~\cite{Goodman:2010qn}, see also
Ref.~\cite{Goodman:2010ku} where $G_{Di}$ were labeled $G_\chi$. The
Wilson coefficients in \eqref{eq:lightDM:Lnf5} are thus, for Dirac
fermion DM, given by

\begin{align}
\hat\C_{1,q}^{(5)} &= \frac{8\pi^2}{e}\,M\,,
& \hat\C_{2,q}^{(5)} &= -i\frac{8\pi^2}{e}\,D\,,\\
	\hat\C_{1,q}^{(6)} &= \left[M_{*}^{\{\text{D5},q\}}\right]^{-2}\,,
& 	\hat\C_{2,q}^{(6)} &= \left[M_{*}^{\{\text{D6},q\}}\right]^{-2}\,,\\
	\hat\C_{3,q}^{(6)} &= \left[M_{*}^{\{\text{D7},q\}}\right]^{-2}\,,
& 	\hat\C_{4,q}^{(6)} &= \left[M_{*}^{\{\text{D8},q\}}\right]^{-2}\,,\\
	\hat\C_{1}^{(7)} &= 3\pi\,\left[M_{*}^{\{\text{D11}\}}\right]^{-3}\,,
& 	\hat\C_{2}^{(7)} &= 3\pi\,\left[M_{*}^{\{\text{D12}\}}\right]^{-3}\,,\\
	\hat\C_{3}^{(7)} &= 2\pi i\,\left[M_{*}^{\{\text{D13}\}}\right]^{-3}\,,
&	\hat\C_{4}^{(7)} &= -2\pi i\,\left[M_{*}^{\{\text{D14}\}}\right]^{-3}\,,\\
	\hat\C_{5,q}^{(7)} &= \left[M_{*}^{\{\text{D1,q}\}}\right]^{-3}\,,
&	\hat\C_{6,q}^{(7)} &= \left[M_{*}^{\{\text{D2,q}\}}\right]^{-3}\,,\\
	\hat\C_{7,q}^{(7)} &= \left[M_{*}^{\{\text{D3,q}\}}\right]^{-3}\,,
&	\hat\C_{8,q}^{(7)} &= -\left[M_{*}^{\{\text{D4,q}\}}\right]^{-3}\,,\\
	\hat\C_{9,q}^{(7)} 	&= \frac{1}{m_q}\left[M_{*}^{\{\text{D9,q}\}}\right]^{-2}\,,
&	\hat\C_{10,q}^{(7)} &= \frac{1}{m_q}\left[M_{*}^{\{\text{D10,q}\}}\right]^{-2}\,.
\label{eq:convert-tensors}
\end{align}
The notation we use above is that the Wilson coefficient
$G_{\chi,Di}$, multiplying the operator $\Q_{Di}$, depends on
$M_*^{\{Di\}}$. That is, in order for only the operator $\Q_{Dj}$ to
contribute one needs to set $M_*^{\{Dj\}}$ to the desired finite
value, while taking $M_*^{\{Di\}}\to \infty$ for $i\ne j$ (and setting
$D=M=0$). In the notation of Ref.~\cite{Goodman:2010qn} the
superscripts on $M_*^{\{Di\}}$ were suppressed.  Note that the above
transformation between the two bases involves complex phases,
signaling that some the operators in~\cite{Goodman:2010qn} are not
Hermitian, but anti-Hermitian. Consequently, the corresponding
parameters $D$, $M_*^{\{D13\}}$, and $M_*^{\{D14\}}$ need to be chosen
purely imaginary.

\paragraph*{Majorana DM.} Similarly, the translation from our basis
defined in
Eqs.~(\ref{eq:dim6:Q2Q4:majorana})-(\ref{eq:dim7:Q7Q8:majorana}) to
that of~\cite{Goodman:2010qn} is given by
\begin{align}
 	\hat\C_{2,q}^{(6)} &= \left[M_{*}^{\{\text{M5},q\}}\right]^{-2}\,,
& 	\hat\C_{4,q}^{(6)} &= \left[M_{*}^{\{\text{M6},q\}}\right]^{-2}\,,\\
\hat\C_{1}^{(7)} &= 3\pi\,\left[M_{*}^{\{\text{M7}\}}\right]^{-3}\,,
& 	\hat\C_{2}^{(7)} &= 3\pi\,\left[M_{*}^{\{\text{M8}\}}\right]^{-3}\,,\\
\hat\C_{3}^{(7)} &= 2\pi i\,\left[M_{*}^{\{\text{M9}\}}\right]^{-3}\,,
&	\hat\C_{4}^{(7)} &= -2\pi i\,\left[M_{*}^{\{\text{M10}\}}\right]^{-3}\,,\\
\hat\C_{5,q}^{(7)} &= \left[M_{*}^{\{\text{M1,q}\}}\right]^{-3}\,,
&	\hat\C_{6,q}^{(7)} &= \left[M_{*}^{\{\text{M2,q}\}}\right]^{-3}\,,\\
\hat\C_{7,q}^{(7)} &= \left[M_{*}^{\{\text{M3,q}\}}\right]^{-3}\,,
&	\hat\C_{8,q}^{(7)} &= -\left[M_{*}^{\{\text{M4,q}\}}\right]^{-3}\,.
\end{align}

For {\em complex scalar DM} the interaction Lagrangian in the basis of
Ref.~\cite{Goodman:2010qn} is given by
\begin{equation}
\label{eq:Goodman:Complex}
{\cal L}_\varphi=\sum_{i} G_{Ci} \Q_{Ci},
\end{equation}
with the operators $\Q_{Ci}$, $i=1,\ldots,6$, (the Wilson coefficients
$G_{Ci}$) are listed in the 2nd (3rd) column of Table~II, right, in
Ref.~\cite{Goodman:2010qn}. The translation of the Wilson coefficients
to our basis for complex scalar DM is thus,
\begin{align}
\label{eq:C1q:C2q:scalar}
\hat \C_{1,q}^{(6)} &= -\frac{i}{2}\left[M_{*}^{\{\text{C3},q\}}\right]^{-2}\,,
& \hat \C_{2,q}^{(6)} &= -\frac{i}{2}\left[M_{*}^{\{\text{C4},q\}}\right]^{-2}\,,
\\
\hat \C_{3,q}^{(6)} &= \left[M_{*}^{\{\text{C1},q\}}\right]^{-2}\,,
&\hat \C_{4,q}^{(6)} &= \left[M_{*}^{\{\text{C2},q\}}\right]^{-2}-\left[M_{*}^{\{\text{C4},q\}}\right]^{-2}\,,\\
\label{eq:C5:C6:scalar}
\hat \C_{5}^{(6)} &= 3\pi\,\left[M_{*}^{\{\text{C5}\}}\right]^{-2}\,,
& \hat \C_{6}^{(6)} &= 2\pi i \,\left[M_{*}^{\{\text{C6}\}}\right]^{-2} + \sum_q \left[M_{*}^{\{\text{C4},q\}}\right]^{-2}\,,
\end{align}
where for clarity we display explicitly the operator dependence of
each $M_*$.  In deriving the above relations we used equation of
motion for the vector current, $\partial_\mu \bar q \gamma^\mu q=0$,
and the relation between the chiral QCD anomaly and the axial current,
valid for each quark flavor separately,
\begin{equation}
\partial_\mu \bar q \gamma^\mu \gamma_5 q= 2 m_q \bar qi\gamma_5 q-\frac{\alpha_s}{4\pi}G^a_{\mu\nu}\tilde G^{a,\mu\nu}\,.
\end{equation}
Note that there is no choice of the scale $M_{*}^{\{\text{C4},q\}}$
that makes the Lagrangian Eq.~\eqref{eq:Goodman:Complex} Hermitian.

\paragraph*{Real scalar DM.} Finally, the translation for the Wilson
coefficients of the real scalar operator basis is given by,
\begin{align}
\hat \C_{3,q}^{(6)} &= \left[M_{*}^{\{\text{R1},q\}}\right]^{-2}\,,
&\hat \C_{4,q}^{(6)} &= \left[M_{*}^{\{\text{R2},q\}}\right]^{-2}\,,\\
\label{eq:C5:C6:scalar:real}
\hat \C_{5}^{(6)} &= 3\pi\,\left[M_{*}^{\{\text{R3}\}}\right]^{-2}\,,
& \hat \C_{6}^{(6)} &= 2\pi i \,\left[M_{*}^{\{\text{R4}\}}\right]^{-2}\,.
\end{align}

In \ddm{} the user can directly input the Wilson coefficients in the basis of~\cite{Goodman:2010qn} by setting the value of $M_*$.
To do this, we provide a function, \verb|SetCoeffMstar|, which takes the following arguments
\begin{Verbatim}[frame=single,xleftmargin=1cm,xrightmargin=1cm,commandchars=\\\{\}]
  SetCoeffMstar["basis",QN[f],value]
\end{Verbatim}
where \verb|QN| $\in \{\texttt{D1,\dots,D16; M1,\dots,M10; C1,\dots, C6; R1,\dots,R4}\}$
and \verb|"basis"| and \verb|f| are the basis and the quark flavor respectively --
see the documentation of \verb|SetCoeff| in Sec.~\ref{sec:program} for further detail.

\bibliography{paper_directDM}

\end{document}